\documentclass[12pt,preprint]{aastex}
\usepackage{epsf}

\newcommand{\be}{\begin{displaymath}}
\newcommand{\ee}{\end{displaymath}}
\newcommand{\bea}{\begin{eqnarray}}
\newcommand{\eea}{\end{eqnarray}}

\newcommand{\kelv}{\ensuremath{\,\mathrm K}}
\newcommand{\kap}[1]{Section \ref{#1}}
\newcommand{\mem}[1]{\ensuremath{\mathrm{ #1}}}
\newcommand{\natlog}[2]{\ensuremath{#1\times 10^{#2}}} 
\newcommand{\msun}{\ensuremath{\, {\rm M}_\odot}}
\newcommand{\jahre}{\ensuremath{\, \mathrm{yr}}}
\newcommand{\abb}[1]{Fig.\,\ref{#1}}
\newcommand{\czw}{\ensuremath{^{12}\mem{C}}}
\newcommand{\ose}{\ensuremath{^{16}\mem{O}}}

\usepackage{natbib}
\usepackage{hyperref}

\shortauthors{Denissenkov et al.}
\shorttitle{MESA Models of Classical Nova Outbursts}

\begin{document}

\title{MESA MODELS OF CLASSICAL NOVA OUTBURSTS: \\ THE MULTICYCLE EVOLUTION AND EFFECTS OF CONVECTIVE BOUNDARY MIXING}


\author{Pavel A. Denissenkov\altaffilmark{1,2,3}, Falk Herwig\altaffilmark{1,3,5}, 
        Lars Bildsten\altaffilmark{4}, \& Bill Paxton\altaffilmark{4}}
\altaffiltext{1}{Department of Physics \& Astronomy, University of Victoria,
       P.O.~Box 3055, Victoria, B.C., V8W~3P6, Canada, 
       pavelden@uvic.ca, fherwig@uvic.ca} 
\altaffiltext{2}{TRIUMF, 4004 Wesbrook Mall, Vancouver, BC V6T~2A3, Canada}
\altaffiltext{3}{Joint Institute for Nuclear Astrophysics}
\altaffiltext{4}{Kavli Institute for Theoretical Physics and Department of Physics, Kohn Hall,
       University of California, Santa Barbara, CA 93106, USA,
       bildsten@kitp.ucsb.edu, paxton@kitp.ucsb.edu}
\altaffiltext{5}{Turbulence in Stellar Astrophysics Program, New Mexico
  Consortium, Los Alamos, NM 87544, USA}

\begin{abstract}
  Novae are cataclysmic variables driven by accretion of H-rich material onto
  a white-dwarf (WD) star from its low-mass main-sequence binary companion.
  New time-domain observational capabilities, such
  as the Palomar Transient Factory and Pan-STARRS, have revealed a diversity of
  their behaviour that should be theoretically addressed. Nova outbursts
  depend sensitively on nuclear physics data, and more readily
  available nova simulations are needed in order to effectively
  prioritize experimental effort in nuclear astrophysics.  In this
  paper we use the MESA stellar evolution code to construct multicycle
  nova evolution sequences with CO WD cores. We explore a range of WD
  masses and accretion rates as well as the effect of different
  cooling times before the onset of accretion. In addition, we study
  the dependence on the elemental abundance distribution of accreted
  material and convective boundary mixing at the core-envelope
  interface. Models with such convective boundary mixing display an
  enrichment of the accreted envelope with C and O from the underlying
  white dwarf that is commensurate with observations. We compare our
  results with the previous work and investigate a new scenario for novae
  with the $^3$He-triggered convection.
\end{abstract} 


\section{Introduction}
\label{sec:intro}
Classical novae are the
result of thermonuclear explosions of H-rich material occurring on
the surfaces of white dwarfs (WDs). Such events typically occur in a close
binary system containing a cold WD primary component and a low-mass
main-sequence (MS) star, the latter filling its Roche lobe
\citep{jh07a,gea98}. The solar-composition material from the envelope of the
secondary component streams through the inner Lagrangian point
to form an accretion disk and eventually,
after having lost its orbital angular momentum, joins the WD.
For sufficiently low accretion rates ($\dot{M}\approx
10^{-11}$\,--\,$10^{-9}\,M_\odot/\mbox{yr}$), the accreted material
accumulates in a thin layer atop the WD until its
base temperature, that rises because of gravitational compression,
reaches a value at which H begins to burn, initially in the pp-chain
reactions and then in the CNO cycle.  
As a result, a thermonuclear runaway (TNR)
ensues causing rapid increases in both the temperature and energy
output.  Peak temperatures during classical nova outbursts can be as
high as $T_{\rm max} = 2 - 4\times10^{8}\kelv$, approaching the virial temperature.  Under such conditions, the
proton-capture reactions of the CNO cycle become so fast that they
build up large amounts of $\beta^+$-unstable isotopes of N, O, and F.
Their temperature-independent decay rates then limit the energy
generation in the CNO-cycle (\citealt{sea78}).  
For neon novae, the activation of the NeNa and MgAl cycles driven by injection of
Ne-seed nuclei from the ONe-rich substrate can lead to nuclear processing of even higher mass isotopes.
After a time period of $10^2-10^3$ seconds,
the electron-degenerate conditions are lifted as the hot convective
envelope expands. Mass loss ensues either from a wind or Roche lobe
overflow triggered by the radius expansion of the burning WD.

Despite much progress over the past decades (e.g. \citealt{jh07a}),
some key aspects of classical novae are still
poorly understood. An important one is the mixing between the
accreted envelope and WD, which is required to explain the
enrichment observed in most nova ejecta in heavy elements, such as C, N, O, and
Ne. These can reach combined total ejected mass fractions of
30\,--\,40\% (e.g., \citealt{gea98}). The TNR peak
temperatures and durations do not allow for fuelling production, so the only possible explanation of this
enrichment is that it originates from the underlying CO or ONe
WDs. The proposed mixing models can be divided into two groups. Either
it is assumed that the envelope and WD get mixed in a thin layer close
to their interface before the TNR ensues (e.g.,
\citealt{pk84,md83,aea04}), or mixing is assumed to be the result of
hydrodynamic boundary mixing at the bottom of the convection zone
triggered by the TNR itself (e.g.,
\citealt{gl95,gea97,gea05,gea07,cea10,cea11a,cea11b}).

The mechanism of mass loss by novae is also not fully
understood.  Apparently, a supersonic outflow, a wind driven by the
super-Eddington luminosity (\citealt{kh94}), and an expansion of
nova ejecta as a result of its common-envelope interaction with the
companion star (e.g., \citealt{lea90}) all contribute to the mass
ejection. 
Finally, there are a few key nuclear reactions relevant for nova nucleosynthesis whose rates still
remain uncertain, e.g. $^{18}$F(p,$\alpha$), $^{25}$Al(p,$\gamma$), and $^{30}$P(p,$\gamma$),
according to \cite{jh07a}.

Nova outbursts are recurrent events, where
depending on the parameters of a particular system, $10^3$\,--\,$10^4$
H-shell flashes would occur in a single binary
system. However, most computer simulations of nova follow only an
individual outburst. An exception are the models by the Tel Aviv group
\citep{pk95,yea05,eea07}, who followed more than 1000 flashes. They adopted mixing of the
first type mentioned above, i.e.\  diffusive
interface mixing during quiet accretion phases between consecutive
outbursts.

In this paper we address and investigate some of these
issues. We describe the simulation assumptions
in \kap{sec:method}.  As a start, we have constructed a grid of multicycle
simulations of CO novae with the masses $0.65\,M_\odot$,
$0.85\,M_\odot$, $1.0\,M_\odot$, $1.15\,M_\odot$, and $1.2\,M_\odot$ that accrete the
solar-composition and CO-enriched material at rates $\dot{M} =
10^{-11}$, $10^{-10}$, and $10^{-9}\,M_\odot/\mbox{yr}$
(\kap{sec:MESA}). The exact value of the WD mass dividing CO and
ONe novae is not well settled. Current estimates suggest a value of about $1.1\,M_\odot$ 
when the effect of binarity is taken into account\citep{gpea03}.
As a verification exercise, we confirm the
result obtained by \cite{gt09} that, for massive and slowly
accreting CO WDs, the peak temperature achieved during the TNR becomes
a much steeper function of WD's central temperature, when the latter
is lower than $10^7\kelv$ (\kap{sec:comp}).
We then investigate, for the
$1.2\,M_\odot$ case, the effect of convective boundary mixing (CBM) during
the TNR (\kap{sec:standmix}).  For these simulations we invoke an
exponentially decaying mixing efficiency model that has already been studied
previously in related stellar evolution phases, such as the
He-shell flash convection zones in AGB stars
\citep[e.g.][]{herwig:99}.  Finally, we report on a new scenario for
novae with the $^3$He-triggered convection (\kap{sec:he3mix}), and
give conclusions in \kap{sec:concl}.

\section{Simulation Assumptions}
\label{sec:method}
The calculations are performed with the stellar evolution code
\emph{Modules for Experiments in Stellar Evolution}
\citep[MESA,][]{pea11}\footnote{MESA web page with detailed instructions: \texttt{http://mesa.sourceforge.net}.}, 
\texttt[rev.\,3611].
From the available options we use the 2005 update of the
OPAL EOS tables (\citealt{rn02}) supplemented for lower
temperatures and densities by the SCVH EOS that includes partial dissociation and ionization 
caused by pressure and temperature (\citealt{sea95}).
Additionally, the HELM (\citealt{ts00}) and PC (\citealt{pc10}) EOSs
are used to cover the regions where the first two EOSs are not
applicable. 
In particular, the HELM EOS takes into account electron-positron pairs
at high temperature, while the PC EOS incorporates crystallization at low temperature. Both assume complete
ionization. There are smooth transitions
between the four EOS tables. When creating WD models and following their subsequent nova evolution,
different parts of our computed stellar models are entering the $\rho$\,--\,$T$ domains covered by all of the above EOSs.
We use the OPAL opacities
(\citealt{ir93,ir96}) supplemented by the low $T$ opacities of
\cite{fergea05}, and by the electron conduction opacities of
\cite{cea07} (for details, see \citealt{pea11}).

As in any other phase of stellar evolution, the adopted nuclear
network has to be as large as necessary in order to account for the
energy generation, yet as small as possible in order to make
computations not too expensive. In MESA, we use the option to solve
the nuclear reaction network, the structure equation and mixing
operators simultaneously. This leads to a more stable numerical behaviour,
however it makes adding species to the nuclear network relatively more
expensive compared to using a nuclear reaction operator split option.

We have started our CO nova simulations with 77 isotopes from H to $^{40}$Ca coupled by 442 reactions
and then gradually reduced these numbers checking that this does not lead to noticeable changes in
the time variation of the peak temperature. As a result,
an acceptable compromise has been found empirically, and
the following 33 isotopes were selected: $^1$H, $^3$He, $^4$He, $^7$Li,
$^7$Be, $^8$B, $^{11}$B, $^{12}$C, $^{13}$C, $^{13}$N, $^{14}$N,
$^{15}$N, $^{14}$O, $^{15}$O, $^{16}$O, $^{17}$O, $^{18}$O, $^{17}$F,
$^{18}$F, $^{19}$F, $^{18}$Ne, $^{19}$Ne, $^{20}$Ne, $^{21}$Ne,
$^{22}$Ne, $^{21}$Na, $^{22}$Na, $^{23}$Na, $^{22}$Mg, $^{23}$Mg,
$^{24}$Mg, $^{25}$Mg, and $^{26}$Mg. These are coupled by 65
reactions, including those of the pp chains (the pep reaction, whose
importance was emphasized by \cite{sea09}, has also been added), CNO
and NeNa cycles. The underlying CO WD models have been prepared using
the same isotopes, while the reaction list was extended to take into
account He and C burning. By default, MESA uses reaction rates from \cite{cf88} and
\cite{aea99}, with preference given to the second source (NACRE). It
includes updates to the NACRE rates for $^{14}$N(p,$\gamma)^{15}$O
(\citealt{iea05}), the triple-$\alpha$ reaction (\citealt{fea05}),
$^{14}$N$(\alpha,\gamma)^{18}$F (\citealt{gea00}), and
$^{12}$C$(\alpha,\gamma)^{16}$O (\citealt{kea02}). Although the main nuclear
path for a classical nova is driven by p-capture reactions and $\beta$-decays,
the $\alpha$-reactions are important for establishing the chemical composition
of its underlying WD. As a test, we have also tried the MESA second option for choosing 
the reactions rates that gives preference to the JINA REACLIB
\citep{jina}. This database has many nuclear reaction rates in common with those
considered by \cite{iea10}. We have not found any significant differences
in our nova $T_{\rm max}(t)$ profiles between the two options, therefore
we decided to stick to the default one.

The options that specify the physics and numerics assumptions of a
MESA simulation are set in an \texttt{inlist} file. We have
started from inlists of two MESA test suite cases relevant for our
problem:
  \begin{itemize}
  \item \texttt{make\_co\_wd} combines
    some ``stellar engineering'' tricks into a procedure that creates CO WD models from a
    range of initial masses (see below),
  \item \texttt{wd2} demonstrates the use of parameters that
    control accretion, as well as mass ejection options relevant for
    nova calculations.
  \end{itemize}

  Test suite case \texttt{make\_co\_wd} uses limits on the opacity
  during the AGB evolution to achieve rapid envelope removal without
  having to compute the details of many thermal pulses which
  eventually culminate in the so-called superwind phase. The
  simulation of this final tip-AGB phase of evolution is not
  straightforward and beyond the scope of this investigation. The
  procedure followed here is different from that described by
  \cite{ww94}, but has in the end the same effect.

  First, a sufficiently massive initial model has to be chosen, e.g. a
  $6\,M_\odot$ pre-MS star for the $0.85\,M_\odot$ WD.  Its evolution
  is computed until the mass of its He-exhausted core reaches a value
  close to the final WD's mass. This phase is shown with the solid blue
  curves in \abb{fig:f1}.  After that, the maximum opacity is
  reduced to a small value (the red curves), the total mass of the
  star is relaxed to $M_{\rm WD}$ (the solid black and dashed green curves), the
  maximum opacity is restored (the small red circles), and finally the
  WD model is given time to relax (the dot-dashed cyan curves).  As a
  result, the stellar model arrives at the WD cooling track, from
  which a WD model with a required central temperature, $T_{\rm WD}$, (a lower $T_{\rm WD}$
  corresponds to a longer WD's cooling time) can be selected (the
  dotted magenta curves).

In reality, a CO or an ONe WD recently formed in a binary system should be surrounded
by a buffer zone of unburnt material (He-rich or CO-rich, respectively), and
quite a large number of nova outbursts have to occur before it will be removed (\citealt{jea03}).
Our WD making procedure does have a step on which the WD models possess such buffer zones.
To avoid a discussion of effects to which the presence of He-rich buffer zones can lead,
we remove them artificially and use naked CO WDs as the initial models for our nova simulations in this paper.

  A set of the more massive CO WD models was obtained differently, by
  letting the $1.0\,M_\odot$ CO WD accrete its core composition
  material until the total masses of $1.15\,M_\odot$ and $1.2\,M_\odot$ were accumulated, after
  which the stars were cooled off to generate the massive CO WD
  models for a range of initial core temperature (Table~\ref{tab:tab1}).
  This special procedure has been employed because it is impossible to obtain a CO WD with
  $M_{\rm WD} \ga 1.0\,M_\odot$ following the evolution of a massive AGB star, unless one artificially
  turns off carbon burning.
  For example, \cite{gpea03} have found that for $M_{\rm ZAMS}$ increasing from $9.3\,M_\odot$
  to $11\,M_\odot$ the final WD core mass for the binary star evolution is changing from
  $1.07\,M_\odot$ to $1.22\,M_\odot$ but, because of carbon burning in the core, it actually
  contains an ONe WD surrounded by a CO buffer zone, the relative mass of the latter
  decreasing from 7\% to 0.8\% for the given mass interval. Therefore, our massive CO WD models
  should be considered as substitutes for such ONe WDs with CO surface buffer zones. Besides, they are used
  for comparison with other nova simulations that involved CO WDs of similar masses.

\section{MESA Models of Multicycle CO Nova Outbursts}
\label{sec:MESA}

For either CO or ONe WD the main properties of its nova
outburst (such as total accreted and ejected masses, peak temperature,
maximum luminosity, envelope expansion velocity, and chemical
composition of the ejecta) depend mainly on the
following four parameters: the WD mass $M_{\rm WD}$, its central
temperature $T_{\rm WD}$ (or luminosity), the accretion rate
$\dot{M}$, and the metallicity of the accreted material
(e.g., \citealt{pk95,tb04,jea07}). In this
paper, we consider only CO WDs in binary systems with solar-metallicity companions, and in this section we
ignore any mixing between the WD and its accreted envelope.

\abb{fig:f2} shows a snapshot of the evolution of the hottest of
our $1.15\,M_\odot$ CO WD models ($T_{\rm WD}\approx
\natlog{3}{7}\kelv$; for the correspondence between the WD's initial central temperature
$T_{\rm WD}$ and luminosity $L_{\rm WD}$, see Table~\ref{tab:tab1}) accreting the solar-composition material with the
rate $\dot{M} = 10^{-10}\,M_\odot$/yr.  The total mass $M = M_{\rm WD}
+ M_{\rm acc}$ increases during an accretion phase and decreases
during a mass-loss event.  The abrupt changes in the abundances at $x_{\rm WD}\approx
-4.6$ and $x_{\rm CE}\approx -8.8$ mark the WD's surface and upper
boundary of the convective envelope, respectively. During an accretion
phase, $x_{\rm WD}$ moves to the left, which signifies that the mass
(of the envelope) to the right of $x_{\rm WD}$ 
increases; $x_{\rm WD}$ shifts to the right during 
a mass-loss event.\footnote{A movie demonstrating the multicycle evolution of the $1.15\,M_\odot$
  CO nova model is available at
  \texttt{http://astro.triumf.ca/nova-movies}.} 

\abb{fig:f2} corresponds to the moment, when the
temperature at the interface between the WD and
accreted envelope has reached its maximum value (a sharp peak on the solid red curve in
the upper-right panel), and
most of the initially abundant $^{12}$C, $^{13}$C, $^{14}$N, and
$^{16}$O nuclei have been transformed into the
$\beta^+$-unstable p-capture product isotopes 
$^{13}$N, $^{14}$O, $^{15}$O, and $^{17}$F (the dot-dashed blue, dashed black, dotted red and
solid magenta curves in the middle-right panel) in the
convective envelope (a region where $\nabla_{\rm rad} > \nabla \ga \nabla_{\rm ad}$ in the lower-right panel).
The $\beta$-decay times of these isotopes limit the energy production,
unless additional C and O is mixed from the layers below the
convection zone.

For the multicycle simulations mass loss is triggered when the models
reach super-Eddington luminosities according to the following
prescription: 
\bea \dot{M} = -2\,\eta_{\rm Edd}\,\frac{(L-L_{\rm
    Edd})}{v^2_{\rm esc}},
\label{eq:mdot}
\eea where $v_{\rm esc} = \sqrt{2GM/R}$, and $L_{\rm Edd} = (4\pi
GcM)/\kappa$. Here, $M$, $R$, and $L > L_{\rm Edd}$ are the mass,
radius, and luminosity of the star, while $\kappa$ is the Rosseland
mean opacity at the surface. The scaling factor has been set to
$\eta_{\rm Edd}=1$. This prescription simply assumes that the excess of nova luminosity over
the Eddington one determines the rate of change of the mass-loss kinetic energy.

The main results of our $1.15\,M_\odot$ CO nova simulations for the
mass accretion rate $\dot{M} = 10^{-10}\,M_\odot$/yr are presented in
\abb{fig:f3}. Cases for accretion of solar and $30\%$ CO enriched
material, as well as for different WD central temperatures are
shown. The case with CO-enriched accretion material is not realistic
because in the majority of actual CO novae the donor star is a Pop I
MS dwarf providing material close to solar to be accreted by the
WD. Therefore, the CO enrichment of nova ejecta comes from the
underlying CO WDs and occurs either before or during the
TNR. Nevertheless, the case with CO enhanced accretor material is
frequently considered in the literature 
as an artificial way to mimic the effect of mixing at the core-envelope interface
(e.g., \citealt{jh98,sea98}).

The effect of the initial WD luminosity (or central temperature) on the strength of
the nova explosion has been analyzed before (e.g., \citealt{sea98,yea05,jh07b}).
As expected, the strongest outburst occurs in the case of the coolest
WD (the dotted red curves). Its lower temperature allows the WD to
accumulate a slightly more massive H-rich envelope on a longer timescale (the
lower-right panel) before the TNR is triggered. This leads to a higher
peak temperature, $T_{\rm max}\approx \natlog{2.29}{8}\kelv$, and to a
longer time of envelope's removal by the mass-loss that extends the
nova evolution track towards lower effective temperatures and larger
radii. The corresponding track (dotted red)
reaches a maximum radius $R\approx 3\,R_\odot$.  Evidently, to produce
a self-consistent model, we should not allow the nova to expand far
beyond its Roche lobe radius $R_{\rm RL}$.  For example, if the
$1.15\,M_\odot$ WD has a $0.6\,M_\odot$ MS companion then, for the
latter to fill its Roche lobe and therefore be able to transfer its
mass onto the WD, the binary rotation period has to be nearly 5 hours
(for a circular orbit with a semi-major axis $a=1.8\,R_\odot$). 
In this case, the WD itself will have a Roche
lobe with $R_{\rm RL}\approx 0.8\,R_\odot$.  The MESA stellar
evolution code has an option to limit the growth of a star beyond
its Roche lobe radius by exponentially increasing the mass-loss rate when $R >
R_{\rm RL}$, however we did not implement this option
in these simulations.

The CO-enriched material ignites much earlier than the
solar-composition case because it contains much larger mass fractions
of CO isotopes that serve as catalysts for H burning in the CNO
cycle. As a result of the lower accreted mass, the TNR peak
temperature reaches only $\natlog{1.59}{8}\kelv$ in this case (the dashed blue
curves in the lower panels of \abb{fig:f3}).
Table~\ref{tab:tab1} summarizes the accreted masses, maximum H-burning
luminosities, and peak temperatures as functions of $M_{\rm WD}$,
$T_{\rm WD}$, $L_{\rm WD}$, and $\dot{M}$ obtained in the nova simulations.
The number of grid zones in the envelopes of the MESA nova models
varies between 500 and 1000, depending on the complexity and evolutionary phase of
the model. A comparable number of grid zones is allocated to the underlying WD.

\section{Comparison with Other Nova Simulations}
\label{sec:comp}

Ami Glasner provided us with the parameters of his 1D model
of a nova outburst occurring on a $1.147\msun$ CO WD. It has a
central temperature $T_{\rm WD} = \natlog{2.4}{7}\kelv$, accretes
solar-composition material at a rate $\dot{M} = 10^{-10}\msun/\jahre$,
and no interface mixing has been adopted \citep{gea11}.
The temperature and luminosity comparison
of that model with our $1.15\,M_\odot$ CO nova simulation with $T_{\rm
 WD}\approx \natlog{2.5}{7}\kelv$ shows very good agreement, in
particular with respect to the amplitudes (\abb{fig:f4}).
The relative differences between the accreted masses and peak temperatures
for the two models are only 16\% and 6\%, respectively.

As a second test, we check if our nova model agrees with findings
reported recently by \cite{gt09} that the peak temperature achieved
during the TNR becomes a much steeper function of $T_{\rm WD}$ when
the latter is lower than $10^7\kelv$. Their study was motivated
by the result obtained earlier by \cite{tb04}, according to which such
cold WDs should be associated with nova outbursts occurring in
binaries with $\dot{M} < 10^{-10}\msun/\jahre$. To carry out this
test, we let the $1.2\msun$ CO WD model cool down to a central
temperature $T_{\rm WD} = \natlog{3.3}{6}\kelv$, let it accrete the
solar-composition material with the rate $\dot{M} =
10^{-11}\msun/\jahre$, and followed the ensuing nova outburst. This
simulation has been complemented with the ones done for $T_{\rm WD} =
7$, $15$, $20$ and $\natlog{30}{6}\kelv$ and the same $\dot{M}$.  The
resulting relation between $T_{\rm max}$ and $T_{\rm WD}$
is very similar to those plotted by Glasner \&
Truran in their Fig.\,1, and our model therefore confirm their
findings.  Quantitatively, for $M_{\rm WD} = 1.2\msun$, $T_{\rm WD} =
\natlog{3.3}{7}\kelv$, and $\dot{M} = 10^{-11}\msun/\jahre$ our model
accretes $M_{\rm acc} = \natlog{1.42}{-4}\msun$ envelope mass before
TNR ignition and its peak temperature is $T_{\rm max} =
\natlog{3.28}{8}\kelv$, while Glasner \& Truran find $M_{\rm acc} =
1.30\times 10^{-4}\msun$ and $T_{\rm max} = \natlog{3.68}{8}\kelv$ for
their slightly more massive WD ($M_{\rm WD} = 1.25\msun$) with $T_{\rm
  WD} = \natlog{4}{7}\kelv$.  Our data are also in a good agreement
with the estimates of $M_{\rm acc}$ and $T_{\rm WD}$ presented by
\cite{tb04} in their Fig.\,8.

Finally, Fig.~\ref{fig:f5} shows solar-scaled mass-averaged abundances in the expanding envelope of a
last model from our simulations of a CO nova with $M_{\rm WD} = 1.15\,M_\odot$, $T_{\rm WD} = 15\times 10^6$ K,
and $\dot{M} = 2\times 10^{-10}\,M_\odot/\mbox{yr}$. These simulations have used an extended nuclear network
that included 48 isotopes from H to $^{40}$Ca coupled by 120 reactions. The accreted material was assumed
to be a mixture of 50\% solar and 50\%
WD's core compositions. This nova model has parameters similar to those of the model CO5 of \cite{jh98}.
A comparison of our final abundances from Fig.~\ref{fig:f5} with those for the model CO5 presented
by \cite{jh98} in their Fig.~1 shows a very good qualitative agreement.

\section{Effects of Convective Boundary Mixing}
\label{sec:mixing}

\subsection{A Standard Mixing Model}
\label{sec:standmix}

The nova models presented in \kap{sec:MESA} do not reproduce the
observed enrichment of nova ejecta in C, N, O, and other heavy
elements \citep[e.g.][]{gea98} because they do not include the
interface mixing between the accreted H-rich envelope and its
underlying CO WD. Recent two- and three dimensional
nuclear-hydrodynamic simulations of a nova outburst have shown that a
possible mechanism of this mixing are the hydrodynamic instabilities
and shear-flow turbulence induced by steep horizontal velocity
gradients at the bottom of the convection zone of the TNR
\citep{cea11b}. These hydrodynamic processes associated with the
convective boundary lead to convective boundary mixing (CBM) at the
base of the accreted envelope into the outer layers of the WD.
As a result, CO-rich (or ONe-rich) material is dredged-up during the
TNR.

Our nova simulations have been performed with the one-dimensional stellar evolution code MESA.
For one-dimensional CBM calculations, MESA provides
a simple model that treats the time-dependent mixing as a
diffusion process, and that approximates the rate of mixing by
an exponentially decreasing function of a distance from the formal
convective boundary, \bea D_{\rm OV} =
D_0\exp\left(-\frac{2|r-r_0|}{fH_P}\right),
\label{eq:DOV}
\eea where $H_P$ is the pressure scale height, and $D_0$ is
a diffusion coefficient, calculated using a mixing-length theory (MLT), that
describes convective mixing at the radius $r_0$ close to the
boundary. In this model $f$ is a free parameter that is
calibrated for each type of convective boundary either
semi-empirically through observations, or through multi-dimensional hydrodynamic
simulations. 

The MESA CBM model is based on the findings in hydrodynamic models \citep{freytag:96} 
that the velocity field, and along with it the mixing expressed in terms of a diffusion coefficient, 
decays exponentially in the stable layer adjacent to a convective boundary. 
Following these findings, the CBM model extends time-dependent mixing according 
to the MLT diffusion coefficient $D_\mathrm{MLT}$ across the Schwarzschild boundary 
with the diffusion coefficient given by Equation (\ref{eq:DOV}). 
The total diffusion coefficient is therefore $D = D_\mathrm{MLT} + D_\mathrm{OV}$. 
In the CBM model adopted here the energy transport in the convectively stable 
layer is assumed to be due to radiation only.
This CBM model was first introduced in stellar evolution
calculations by \citet{hea97}. This, or very similar models, have been applied to
several related situations in stellar evolution. The most relevant,
because similar, case is CBM at the bottom of the He-shell flash (or
pulse-driven) convection zone (PDCZ) in AGB stars
\citep[e.g.][]{herwig:99,miller-bertolami:06,weiss:09}.  The
consequences include larger \czw\ and \ose\ abundances in the intershell,
in agreement with observations of H-deficient post-AGB stars
if $f_\mathrm{PDCZ}\sim0.008$ \citep{werner:06}. Multidimensional
hydrodynamic simulations of He-shell flash convection
seem to support this value of $f_\mathrm{PDCZ}$ \citep{herwig:06,hea07}, but more
sophisticated numerical hydrodynamics work is needed. 

The MESA CBM model is meant to represent a wide range of hydrodynamic instabilities 
that may contribute to mixing at the convective boundary. The cases considered 
by \cite{freytag:96} featured shallow near-surface convection zones with a small 
ratio of the stability in the unstable and stable zones. These convection zones 
display the classical overshoot picture in which coherent convective systems cross 
the convective boundary and then turn around due to buoyancy effects. 
The boundaries of shell-flash convection, such as those in novae or in AGB stars, 
are much stiffer, and coherent convective blobs cannot cross the convective boundary. 
Instead, shear motion, induced by convective flows and internal gravity waves lead 
to mixing at the convective boundaries in which the Kelvin-Helmholtz instability 
plays an important role \citep{herwig:06,cea11b}.
The amount of CBM, as expressed in the free parameter $f$, depends 
on the details of the specific conditions, including the relative stability of 
the stable to unstable side of the boundary as well as the vigour of the convection. 
While for shallow surface convection $f$ was found to be in the range $0.25 \dots 1.0$,
hydrodynamic and semi-empirical studies show that $f=0.008$ is appropriate 
for the bottom of the He-shell flash convection zone.

For the nova simulations we adopt $f_\mathrm{nova} = 0.004$ at the
bottom of the TNR convective zone, that eventually includes most of
the accreted envelope. 
This number is of the same order of magnitude (within a factor of two)
compared to CBM efficiencies that have been found to reproduce observables
related to the He-shell flash convection zone in AGB stars.
As a result, $^{12}$C and $^{16}$O are mixed
from the WD below, and their combined mass fraction in the convective
zone reaches $Z_{\rm CO}\approx 0.29$. Such a CO abundance
is similar to the average mass fraction of CNO elements observed in
the ejecta of CO novae. With this increase in the total CO abundance,
the simulation produces a fast CO nova. The surface (bolometric)
luminosity increases by six orders of magnitude on a timescale of 50
seconds (the solid red curve in the upper-right panel in \abb{fig:f6}). The radius increases
on a longer timescale, of the order of $10^4$ seconds (the lower-left
panel). This corresponds to a surface expansion velocity of nearly
300\,km/s. The velocity exceeds the speed of sound only in the outer
layers of the expanding envelope, which has a negligible relative mass
(the lower-right panel).  Given the very short evolution timescale of
this nova model, its post-TNR mass-loss cannot anymore be caused by
the super-Eddington luminosity alone because the associated mass-loss
rate (\ref{eq:mdot}) is too slow.  Instead, the envelope would rather
quickly fill the WD's Roche lobe, after which it would probably be
expelled from the binary system as a result of its (common-envelope)
interaction with the secondary component (e.g., \citealt{lea90}). The
details of this process are not yet understood, and we do therefore not
attempt multicycle simulations of nova models with CBM.

As an additional test, we compare the maximum
hydrogen-burning luminosities $L_{\rm H}$ achieved in the basic
$1.2\msun$ CO nova models with simple estimates based on the
approximation of the nuclear energy generation rate limited by the
mass fraction of CNO elements $Z_{\rm CNO}$ in the convective
envelope of a CO nova during its TNR. When limited by the $\beta$ decays
(e.g. ``Hot'' CNO cycle),
$$   
\varepsilon_{\rm max} \approx 5.6\times 10^{13}\left(\frac{Z_{\rm CNO}}{0.01}\right)\ \ \
\mbox{erg}\cdot\mbox{g}^{-1}\cdot\mbox{s}^{-1},
$$   
\citep{gea07}, and given that $L_{\rm H}\approx
\varepsilon_{\rm max}M_{\rm conv}$, where $M_{\rm conv}$ is the mass
of the convective envelope, the hydrogen-burning luminosity can be
estimated as:
\bea
\log_{10}\frac{L_{\rm H}}{L_\odot}\approx 8.48 + \log_{10}\frac{Z_{\rm CNO}}{0.01} + \log_{10}\frac{M_{\rm conv}}{10^{-5}\msun}.
\label{eq:LH}
\eea 

Without any CBM the CNO abundance is $Z_{\rm CNO} =
Z_\odot = 0.019$ and $M_{\rm conv} \approx M_{\rm acc} = 1.9\times
10^{-5}\msun$ (data from Table~\ref{tab:tab1} for $M_{\rm WD} =
1.2\msun$, $T_{\rm WD} = \natlog{3.0}{7}\kelv$, and $\dot{M} =
10^{-10}\msun/\jahre$), in which case the equation (\ref{eq:LH})
estimates $\log_{10}(L_{\rm H}/L_\odot)\approx 9.04$, while our
numerical simulations give 8.74. In models with CBM the mass of the
convective envelope $M_{\rm conv}\approx 4.2\times 10^{-5}\msun$
exceeds the accreted mass by the amount of material dredged up from
the WD core. In this case, $Z_{\rm CNO}\approx Z_{\rm CO}\approx 0.29$,
which results in $\log_{10}(L_{\rm H}/L_\odot) = 10.56$. The numerical
simulations give $\log_{10}(L_{\rm H}/L_\odot) = 10.67$. The maximum
H-burning temperature reached during the outburst of our mixed CO nova
is $\natlog{2.32}{8}\kelv$ which is $\natlog{7}{6}\kelv$
higher than $T_{\rm max}$ in the corresponding unmixed model.

\subsection{Mixing Caused by $^3$He Burning}
\label{sec:he3mix}

\cite{shb09} have quantified the role of $^3$He in the onset of a
nova. They have shown that if the mass fraction of $^3$He in the
H-rich material accreted onto a WD is higher than
$X(^3\mbox{He})=2\times 10^{-3}$ then convection in the nova envelope
is triggered by the $^3$He($^3$He,2p)$^4$He reaction, rather than by
$^{12}$C(p,$\gamma)^{13}$N.  This alters the amount of mass that is
accreted prior to a nova outburst and should therefore be taken into
account when comparing the observed and theoretically predicted nova
rates. As a likely place for novae with the $^3$He-triggered
convection, \cite{shb09} consider binaries in which a low-mass MS
component has undergone such significant mass-loss that it now exposes
its formerly deep layers where $^3$He was produced in a large amount
as a result of incomplete pp-chain reactions (the so-called ``$^3$He
bump'').

In our simulations, we have found a variation of this $^3$He-triggered
convection scenario, i.e.\ that the nova can generate the $^3$He
{\it in situ}. The new scenario is based on the results obtained by \cite{tb04},
who have demonstrated that WDs accreting with rates $\dot{M} <
10^{-10}\msun/\jahre$ should maintain their central temperatures at
the level of $T_{\rm WD} < 10^7\kelv$. We have presented such a CO WD
model with $M_{\rm WD} = 1.2\msun$, $T_{\rm WD} =
\natlog{3.3}{6}\kelv$ and $\dot{M} = 10^{-11}\msun/\jahre$ in
\kap{sec:comp}. In this model a large amount of $^3$He,
$X(^3\mbox{He})\approx 5\times 10^{-3}$, is produced at the base of
the accreted envelope (the dot-dashed blue curve in the middle panel in
\abb{fig:f7}). The ``sloped $^3$He enhancement'' is formed as a result of
incomplete pp-chain reactions, like the $^3$He bump in low-mass MS
stars.  Eventually, $^3$He ignition triggers convection, as predicted
by \cite{shb09}. When we include CBM in this model, as we did for the
standard mixing model using $f_\mathrm{nova}=0.004$ in the equation
(\ref{eq:DOV}), the $^3$He-driven convection penetrates into the WD's
outer layers and dredges up large amounts of C and O into the
convective envelope (the middle panel in \abb{fig:f8}), allowing
subsequently for a fast nova.  Note that this interface mixing occurs
before the TNR, when the maximum temperature is still lower than
$\natlog{5}{7}\kelv$.  It is not until the $^3$He abundance in the
convective zone decreases below its solar value that the major TNR
driven by H-burning in the CNO cycle will ensue. 

A comparison of the time intervals between successive nova outbursts (the accretion times),
$t_{\rm acc} = M_{\rm acc}/\dot{M}$, that can be estimated using the corresponding numbers from Table~\ref{tab:tab1},
for the $1.2\,M_\odot$ models with $T_{\rm WD} = 20\times 10^6$ K and $T_{\rm WD} = 3.3\times 10^6$ K,
and with the accretion rates $10^{-10}\,M_\odot/\mbox{yr}$ and $10^{-11}\,M_\odot/\mbox{yr}$, respectively,
shows that for one event involving the cold and slowly accreting WD there should be nearly 59 events
occurring on the hot and faster accreting WD, provided that the both types of cataclysmic variables
are already present in equal amounts. Given that the last assumption is not true, because the cooling time
for the second model is much longer than that for the first one, the relative observational frequency
of novae with the $^3$He-triggered convection should actually be very low. We plan to study nova
models with the $^3$He-triggered convection in more detail, even though as a purely theoretical case, in our
future work.

\section{Conclusion}
\label{sec:concl}

We have presented a grid of nova simulations with the one-dimensional stellar
evolution code MESA, and provided some tests and comparison with works by
others, to demonstrate that MESA can generate state-of-the-art nova
simulations. In addition, we have investigated the effect of convective
boundary mixing at the bottom of the TNR convection
zone. Interestingly, the CBM efficiency that reproduces observed CO
enhancements is of the same order of magnitude (within a factor two)
compared to CBM efficiencies that have been found to reproduce
observables related to the He-shell flash convection zone in AGB
stars. We do not consider this a coincidence, but rather it is likely
that in both cases very similar physics of CBM is at play, namely shear motion of
convective flows and internal gravity waves leading to mixing at the convective boundary,
in which the Kelvin-Helmholtz instability plays an important role. It will be
exciting to study both phenomena side by side in the future. For
example, the one-dimensional simulations performed here are related to the results of
multi-dimensional simulations of CBM reported by \cite{hea07}.  The
CBM mixing parameter $f_\mathrm{nova}=0.004$ turns out to be sufficient
to reproduce both the observed heavy-element enrichment of CO novae,
as well as amounts of CO-rich material dredged up in numerical
simulations of the Kelvin-Helmholtz and other hydrodynamic
instabilities in the novae case (e.g.,
\citealt{cea10,cea11a,cea11b}).

Further, we have studied slow accretion of solar-composition
material onto a CO WD with the central temperature $T_{\rm WD} =
\natlog{3.3}{6}\kelv$, exploring the original scenario proposed by
\citet{tb04} in which nova outbursts can actually occur under such
conditions. We have found that incomplete pp-chain reactions lead to
the formation of a sloped $^3$He enhancement at the base of the
accreted envelope in this case, with the maximum mass fraction
$X(^3\mbox{He})\approx 5\times 10^{-3}$.  As predicted by \cite{shb09}
for this high abundance, $^3$He ignites before the major TNR, and this
triggers the development of a convective zone adjacent to the WD's
surface. When complemented with CBM mixing, sufficiently large amounts
of C and O are mixed into the envelope to produce a fast CO nova.
These new results may suggest a more probable scenario, as compared to
the one proposed by \cite{shb09}, for the $^3$He-triggered
novae. Clearly this aspect of the nova evolution deserves further
investigation, in spite of the fact that the relative observational frequency
of novae with the $^3$He-triggered convection is expected to be very low.

\acknowledgements This research has been supported by the National
Science Foundation under grants PHY 11-25915 and AST 11-09174. This
project was also supported by JINA (NSF grant PHY 08-22648) and
TRIUMF. Herwig acknowledges funding from NSERC through a Discovery
Grant. Denissenkov and Herwig have benefited from discussions with
Lothar Buchmann, Barry Davids, Ami Glasner, Chris Ruiz, Alan Shotter,
Jim Truran, and Michael Wiescher.  We would like to thank Ami Glasner for
providing the parameters of his $1.147\msun$ CO nova model for the
comparison with our model.

\begin{thebibliography}{}

\bibitem[Alexakis et al.(2004)]{aea04}
Alexakis, A., Calder, A. C., Heger, A., Brown, E. F., Dursi, L. J., Truran, J. W., 
Rosner, R., Lamb, D. Q., Timmes, F. X., Fryxell, B., Zingale, M., Ricker, P. M., \& Olson, K.~2004, ApJ, 602, 931 

\bibitem[Angulo et al.(1999)]{aea99}
Angulo, C., et al.~1999, Nucl. Phys. A, 656, 3

\bibitem[Casanova et al.(2010)]{cea10}
Casanova, J., Jos\'{e}, J., Garc\'{i}a-Berro, E., Calder, A., \& Shore, S.~N.~2010, A\&A, 513, L5   

\bibitem[Casanova et al.(2011a)]{cea11a}
Casanova, J., Jos\'{e}, J., Garc\'{i}a-Berro, E., Calder, A., \& Shore, S.~N.~2011a, A\&A, 527, A5   

\bibitem[Casanova et al.(2011b)]{cea11b}
Casanova, J., Jos\'{e}, J., Garc\'{i}a-Berro, E., Shore, S.~N., \& Calder, A.~2011b, Nature, 478, 490  

\bibitem[Cassisi et al.(2007)]{cea07}
Cassisi, S., Potekhin, A.~Y., Pietrinferni, A., Catelan, M., \& Salaris, M.~2007, ApJ, 661, 1094

\bibitem[Caughlan, \& Fowler(1988)]{cf88}
Cauglan, G.~R., \& Fowler, W.~A.~1988, At. Data Nucl. Data Tables, 40, 283

\bibitem[Cyburt et al.(2010)]{jina}
Cyburt, R.~H., Amthor, A.~M., Ferguson, R., Meisel, Z., Smith, K., Warren, S., Heger, A., Hoffman, R.~D.,
Rauscher, T., Sakharuk, A., Schatz, H., Thielemann, F.~K., \& Wiescher, M.~2010, ApJS, 189, 240

\bibitem[Epelstain et al.(2007)]{eea07}
Epelstain, N., Yaron, O., Kovetz, A., \& Prialnik, D.~2007, MNRAS, 374, 1449

\bibitem[Ferguson et al.(2005)]{fergea05}
Ferguson, J.~W., Alexander, D.~R., Allard, F., Barman, T., Bodnarik, J.~G., Hauschildt, P.~H.,
Heffner-Wong, A., \& Tamanai, A.~2005, ApJ, 623, 585

\bibitem[Freytag, Ludwig, \& Steffen(1996)]{freytag:96}
Freytag, B., Ludwig, H.-G., \& Steffen, M.~1996, A\&A, 313, 497

\bibitem[Fynbo et al.(2005)]{fea05}
Fynbo, H. O. U., et al.~2005, Nature, 433, 136

\bibitem[Gehrz et al.(1998)]{gea98}
Gehrz, R. D., Truran, J. W., Williams, R. E., \& Starrfield, S.~1998, PASP, 110, 3    

\bibitem[Gil-Pons et al.(2003)]{gpea03}
Gil-Pons, P., Garc\'{i}a-Berro, E., Jos\'{e}, J., Hernanz, M., \& Truran, J.~W.~2003, A\&A, 407, 1021

\bibitem[Glasner, \& Livne(1995)]{gl95}
Glasner, S.~A., \& Livne, E.~1995, ApJ, 445, L149

\bibitem[Glasner, Livne, \& Truran(1997)]{gea97}
Glasner, S.~A., Livne, E., \& Truran, J.~W.~1997, ApJ, 475, 754

\bibitem[Glasner, Livne, \& Truran(2005)]{gea05}
Glasner, S.~A., Livne, E., \& Truran, J.~W.~2005, ApJ, 625, 347 

\bibitem[Glasner, Livne, \& Truran(2007)]{gea07}
Glasner, S.~A., Livne, E., \& Truran, J.~W.~2007, ApJ, 665, 1321

\bibitem[Glasner \& Truran(2009)]{gt09}
Glasner, S.~A., \& Truran, J.~W.~2009, ApJ, 692, L58 

\bibitem[Glasner, Livne, \& Truran(2011)]{gea11}
Glasner, S.~A., Livne, E., \& Truran, J.~W.~2011, arXiv:1111.6777v2 [astro-ph.SR]

\bibitem[G\"{o}rres et al.(2000)]{gea00}
G\"{o}rres, J., Arlandini, C., Giesen, U., Heil, M., K\"{a}ppeler, F., Leiste, H.,
Stech, E., \& Wiescher, M.~2000, Phys. Rev. C, 62, 055801

\bibitem[Herwig et al.(1997)]{hea97}
Herwig, F., Bl\"{o}cker, T., Sch\"{o}nberner, D., \& El Eid, M.~1997, A\&A, 324, L81  

\bibitem[Herwig et al.(1999)]{herwig:99}
Herwig, F., Bl\"{o}cker, T., Langer, N., \& Driebe, T.~1999, A\&A, 349, L5 

\bibitem[Herwig et al.(2006)]{herwig:06}
Herwig, F., Freytag, B., Hueckstaedt, R.~M., \& Timmes, F.~X.~2006, ApJ, 642, 1057 

\bibitem[Herwig et al.(2007)]{hea07}
Herwig, F., Freytag, B., Fuchs, T., Hansen, J. P., Hueckstaedt, R. M., Porter, D. H., 
Timmes, F. X., \& Woodward, P. R.~2007, in Why Galaxies Care About AGB Stars: Their Importance 
as Actors and Probes, eds. F. Kerschbaum, C. Charbonnel, and R. F. Wing, 
ASP Conference Series (ASP: San Francisco), 378, 43

\bibitem[Iglesias \& Rogers(1993)]{ir93}
Iglesias, C.~A., \& Rogers, F.~J.~1993, ApJ, 412, 752

\bibitem[Iglesias \& Rogers(1996)]{ir96}
Iglesias, C.~A., \& Rogers, F.~J.~1996, ApJ, 464, 943

\bibitem[Iliadis et al.(2010)]{iea10}
Iliadis, C., Longland, R., Champagne, A.~E., Coc, A., \& Fitzgerald, R.~2010, Nucl. Phys. A, 841, 31

\bibitem[Imbriani et al.(2005)]{iea05}
Imbriani, G., et al.~2005, Eur. Phys. J. A, 25, 455

\bibitem[Jos\'{e} \& Hernanz(1998)]{jh98}
Jos\'{e}, J., \& Hernanz, M.~1998, ApJ, 494, 680  

\bibitem[Jos\'{e} et al.(2003)]{jea03}
Jos\'{e}, J., Hernanz, M., Garc\'{i}a-Berro, E., \& Gil-Pons, P.~2003, ApJ, 597, L41

\bibitem[Jos\'{e} et al.(2007)]{jea07}
Jos\'{e}, J., Garc\'{i}a-Berro, E., Hernanz, M., \& Gil-Pons, P.~2007, ApJ, 662, L103

\bibitem[Jos\'{e} \& Hernanz(2007a)]{jh07a}
Jos\'{e}, J., \& Hernanz, M.~2007a, J. Phys. G: Nucl. Part. Phys., 34, R431 

\bibitem[Jos\'{e} \& Hernanz(2007b)]{jh07b}
Jos\'{e}, J., \& Hernanz, M.~2007b, M\&PS, 42, 1135                          

\bibitem[Kato \& Hachisu(1994)]{kh94}
Kato, M., \& Hachisu, I.~1994, ApJ, 437, 802 

\bibitem[Kunz et al.(2002)]{kea02}
Kunz, R., Fey, M., Jaeger, M., Mayer, A., Hammer, J.~W., Staudt, G.,
Harissopulos, S., \& Paradellis, T.~2002, ApJ, 567, 643

\bibitem[Livio \& Truran(1987)]{lt87}
Livio, M., \& Truran, J.~W.~1987, ApJ, 318, 316

\bibitem[Livio et al.(1990)]{lea90}
Livio, M., Shankar, A., Burkert, A., \& Truran, J.~W.~1990, ApJ, 356, 250

\bibitem[MacDonald(1983)]{md83}
MacDonald, J.~1983, ApJ, 273, 289

\bibitem[Miller Bertolami et al.(2006)]{miller-bertolami:06}
Miller Bertolami, M.~M., Althaus, L.~G., Serenelli, A.~M., \& Panei, J.~A.~2006, A\&A, 449, 313

\bibitem[Paxton et al.(2011)]{pea11}
Paxton, B., Bildsten, L., Dotter, A., Herwig, F., Lessafre, P., \& Timmes, F.~2011, ApJS, 192, 3

\bibitem[Potekhin \& Chabrier(2010)]{pc10}
Potekhin, A.~Y., \& Chabrier, G.~2010, Contrib. Plasma Phys, 50, 82

\bibitem[Prialnik(1986)]{p86}
Prialnik, D.~1986, ApJ, 310, 222

\bibitem[Prialnik \& Kovetz(1984)]{pk84}
Prialnik, D., \& Kovetz, A.~1984, ApJ, 281, 367

\bibitem[Prialnik \& Kovetz(1995)]{pk95}
Prialnik, D., \& Kovetz, A.~1995, ApJ, 445, 789

\bibitem[Rogers \& Nayfonov(2002)]{rn02}
Rogers, F.~J., \& Nayfonov, A.~2002, ApJ, 576, 1064

\bibitem[Shen \& Bildsten(2009)]{shb09}
Shen, K.~J., \& Bildsten, L.~2009, ApJ, 692, 324

\bibitem[Starrfield, Truran, \& Sparks(1978)]{sea78}
Starrfield, S., Truran, J.~W., \& Sparks, W.~M.~1978, ApJ, 226, 186

\bibitem[Starrfield et al.(1998)]{sea98}
Starrfield, S., Truran, J.~W., Wiescher, M.~C., \& Sparks, W.~M.~1998, MNRAS, 296, 502

\bibitem[Starrfield et al.(2009)]{sea09}
Starrfield, S., Iliadis, C., Hix, W.~R., Timmes, F.~X., \& Sparks, W.~M.~2009, ApJ, 692, 1532

\bibitem[Saumon, Chabrier, \& van Horn(1995)]{sea95}
Saumon, D., Chabrier, G., \& van Horn, H.~M.~1995, ApJS, 99, 713

\bibitem[Timmes \& Swesty(2000)]{ts00}
Timmes, F.~X., \& Swesty, F.~D.~2000, ApJS, 126, 501

\bibitem[Townsley \& Bildsten(2004)]{tb04}
Townsley, D.~M., \& Bildsten, L.~2004, ApJ, 600, 390

\bibitem[Wagenhuber \& Weiss(1994)]{ww94}
Wagenhuber, J., \& Weiss, A.~1994, A\&A, 286, 121

\bibitem[Weiss \& Ferguson(2009)]{weiss:09}
Weiss, A., \& Ferguson, J.~W.~2009, A\&A, 508, 1343

\bibitem[Werner \& Herwig(2006)]{werner:06}
Werner, K., \& Herwig, F.~2006, PASP, 118, 183 

\bibitem[Yaron et al.(2005)]{yea05}
Yaron, O., Prialnik, D., Shara, M.~M., \& Kovetz, A.~2005, ApJ, 623, 398

\end{thebibliography}

\begin{figure}
\epsfxsize=10cm
\epsffile[40 180 350 695]{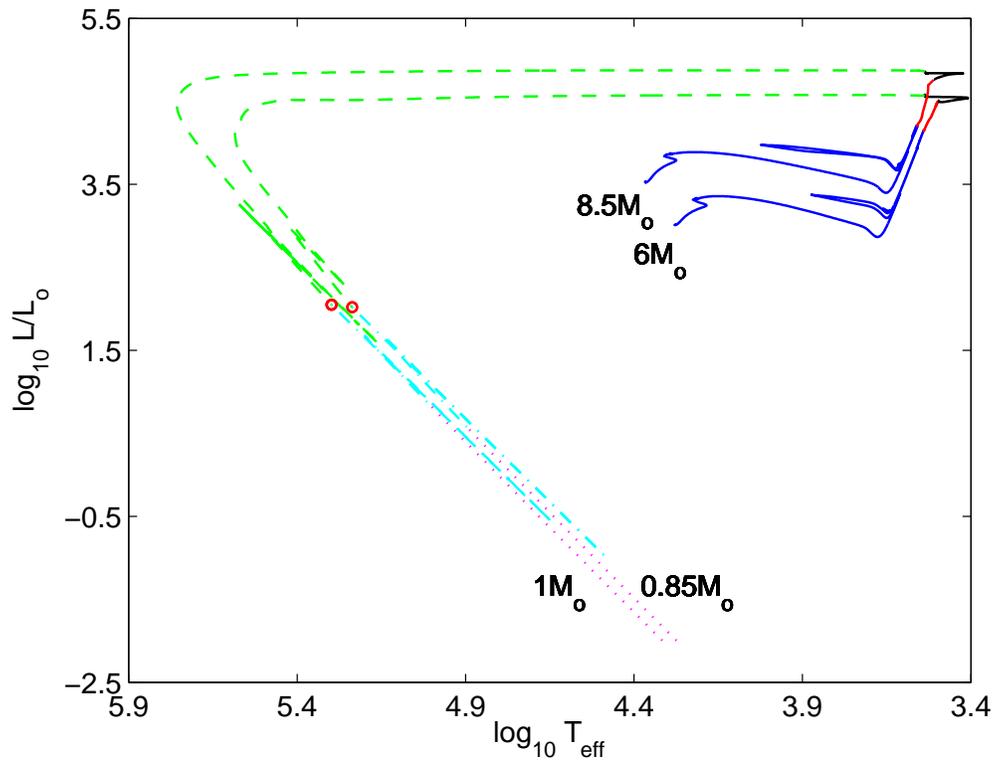}
\caption{Generation of CO WD models with masses $0.85\msun$ and $1\msun$ using MESA and
         ``stellar engineering'' tricks (see text). The computations have used the $6\msun$ and $8.5\msun$
         pre-MS stars as the initial models (the pre-MS evolution is not shown). 
         The detailed stellar evolution computations (the solid blue
         curves) are followed by those in which the maximum opacity is reduced to a small value (the red curves)
         and the total mass of the star is relaxed to $M_{\rm WD}$ (the solid black and dashed green curves).
         After that, the maximum opacity is restored (the small red circles) and the WD model is given time
         to relax (the dot-dashed cyan curves). Finally, the WD model is cooled off to a desired
         temperature (the dotted magenta curves).
         }
\label{fig:f1}
\end{figure}


\begin{figure}
\epsfxsize=10cm
\epsffile [60 210 380 695] {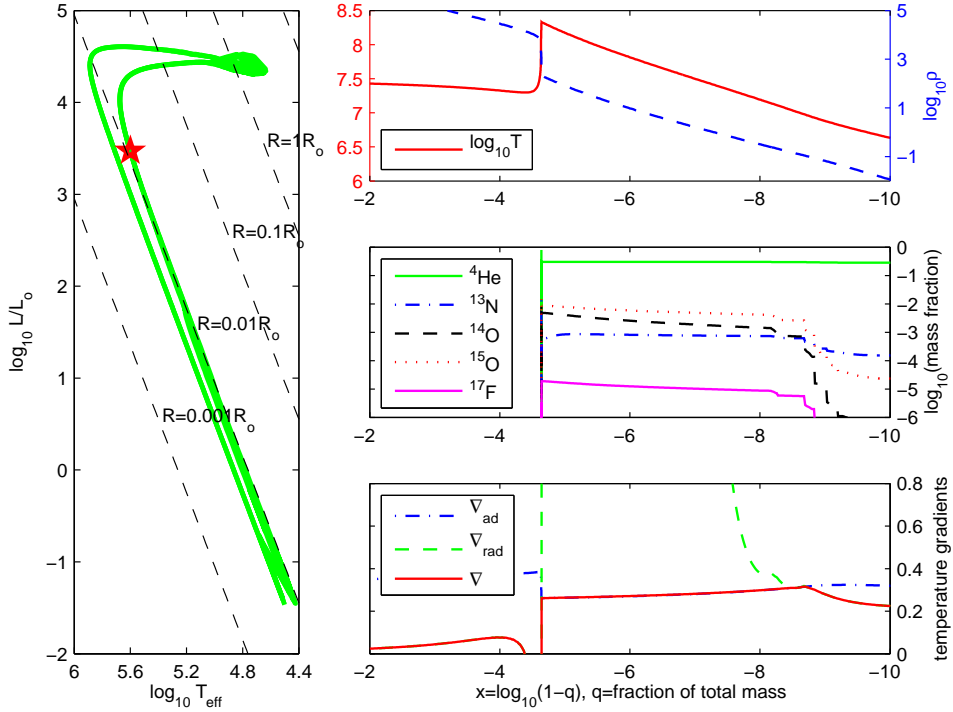}
\caption{ A snapshot of the evolution of the $1.15\msun$ CO WD with
  $T_\mathrm{WD}=\natlog{3}{7}\kelv$, accreting the
  solar-composition material with the rate $10^{-10}\msun/\jahre$.
  Left panel: a track in the Hertzsprung-Russell
  diagram for the whole nova multicycle evolution (accretion, TNR, mass loss).
  The star symbol shows the model internal structure of which is displayed on the right.
  The three right panels depict internal profiles of
  various stellar structure parameters as functions of the quantity
  $x=\log_{10}(1-q)$, where $q = M_r/(M_{\rm WD}+M_{\rm acc})$ is a
  relative mass coordinate.  Here, $M_r$ is the mass inside a sphere
  of the radius $r$, while $M_{\rm acc}$ is the accreted mass. For
  $M_{\rm WD}\approx \msun$, a value $x$ of the abscissa approximates
  to the decimal logarithm of solar masses located to the right of
  this coordinate.  Upper-right: temperature (solid red) and density
  (dashed blue); middle-right: mass fractions of some isotopes; bottom
  right: adiabatic $\nabla_{\rm ad}$, radiative $\nabla_{\rm rad}$,
  and actual $\nabla$ temperature gradients (logarithmic and with
  respect to pressure, dot-dashed blue, dashed green, and solid red
  lines).  In the convectively unstable region, $\nabla_{\rm rad} > \nabla \ga \nabla_{\rm
    ad}$. }
\label{fig:f2}
\end{figure}


\begin{figure}
\epsffile [75 250 480 695] {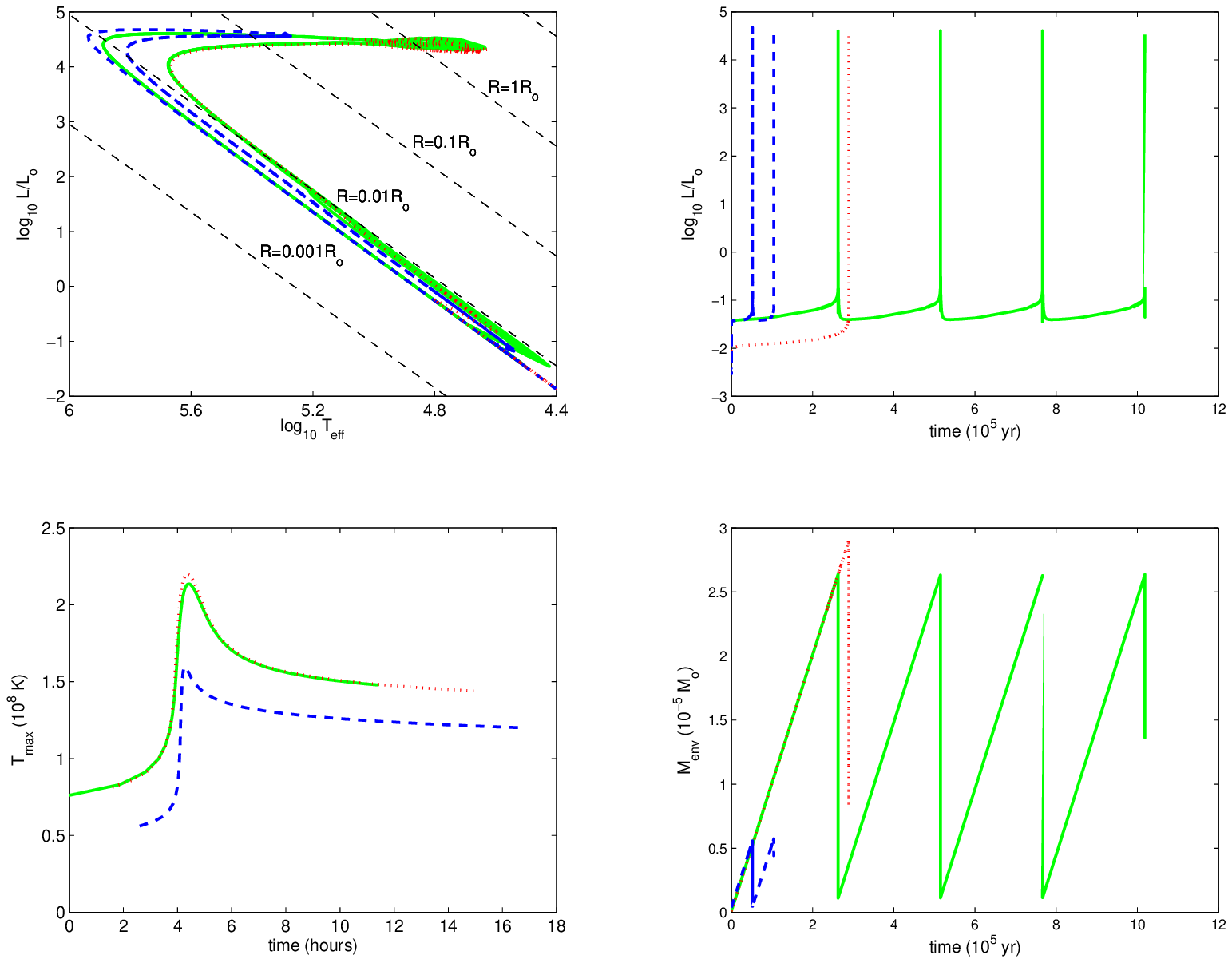}
\caption{A comparison of three nova models that started their
  evolution with the CO WDs having the same mass ($M_{\rm WD} =
  1.15\msun$) and with the same accretion rate ($\dot{M} =
  10^{-10}\msun/\jahre$). The differences in their evolution tracks
  (with dashed black lines of constant radius, upper-left panel), luminosity
  curves (upper-right panel), maximum temperatures of H burning
  (lower-left panel), and accreted envelope masses (lower-right panel)
  are caused by their different WD initial central temperatures
  ($T_{\rm WD} = \natlog{3}{7}\kelv$ for the dashed blue and solid green curves, and
  $T_{\rm WD} = \natlog{1.5}{7}\kelv$ for the dotted red curve) and the
  chemical compositions of accreted material (the solar composition
  for the solid green and dotted red curves, and a mixture of 70\% solar and 30\% CO
  WD compositions for the dashed blue curve).  }
\label{fig:f3}
\end{figure}


\begin{figure}
\epsffile [60 225 480 695] {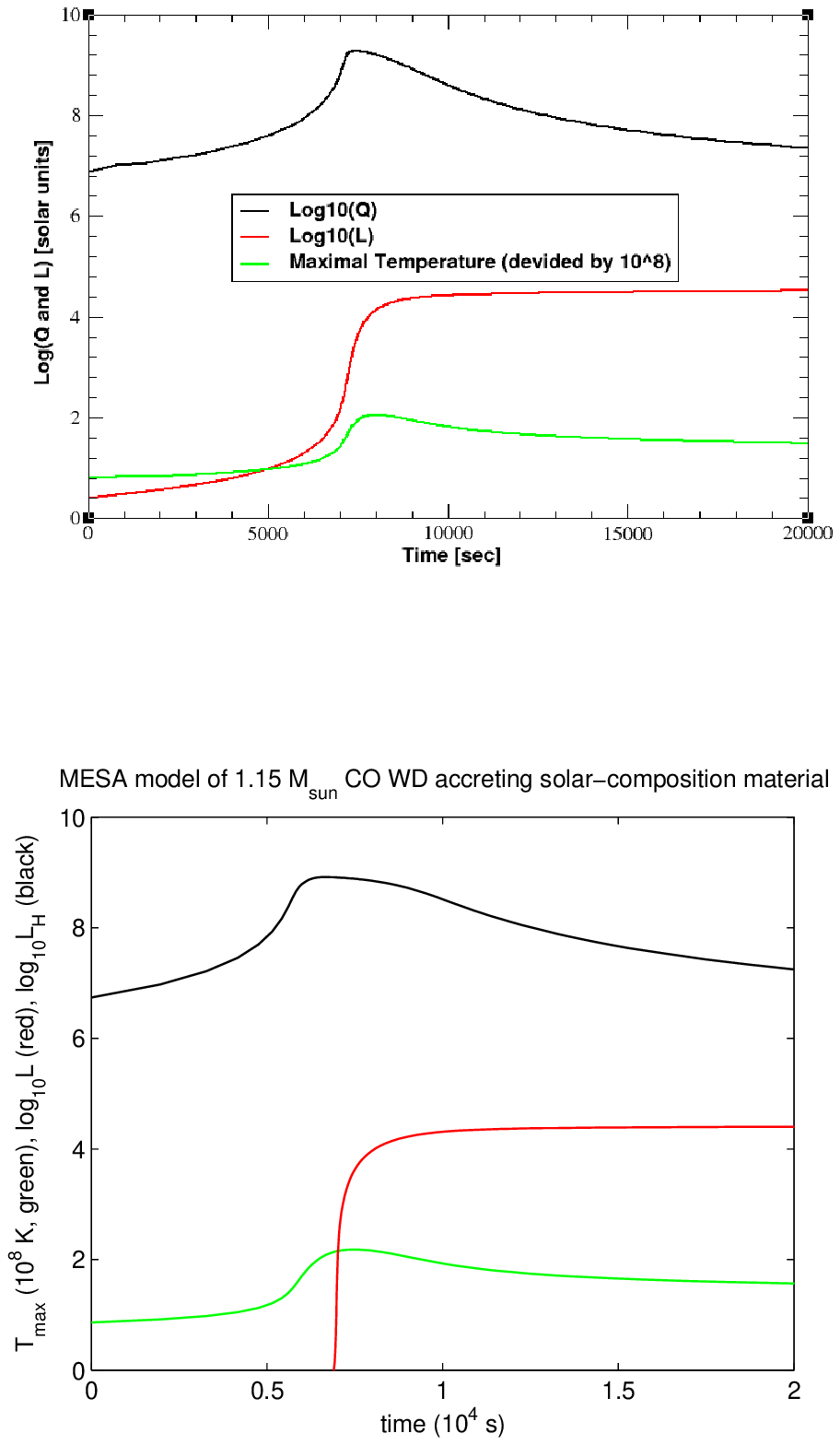}
\caption{A comparison of the hydrogen burning luminosity ($Q$ and
  $L_{\rm H}$, black curves), surface luminosity ($L$, red curves),
  and maximum temperature (green curves) evolution profiles from Ami
  Glasner's $1.147\,M_\odot$ (the upper panel) and our $1.15\,M_\odot$
  (the lower panel) CO nova models accreting solar-composition
  material with the same rate, $\dot{M} = 10^{-10}\msun/\jahre$.
  The difference in the rise of the surface luminosity is probably explained
  by different initial conditions and our better modeling of the evolution
  preceding the TNR.}

\label{fig:f4}
\end{figure}

\begin{figure}
\epsfxsize=10cm
\epsffile[40 180 350 695]{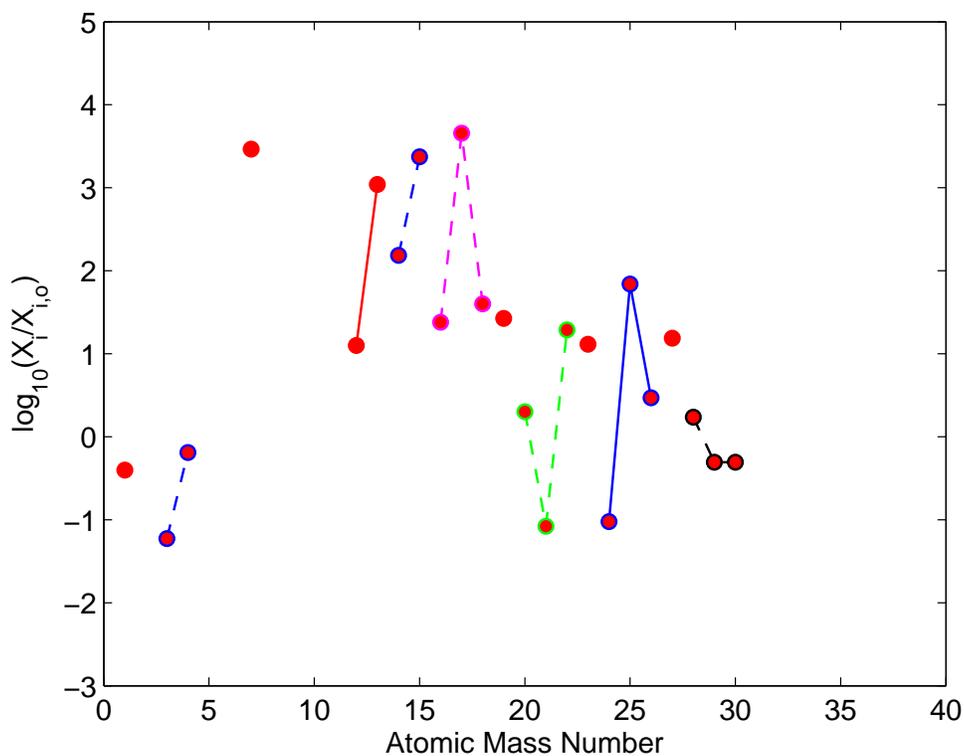}
\caption{Solar-scaled mass-averaged abundances in the expanding envelope of a last model from our
         simulations of a CO nova with $M_{\rm WD} = 1.15\,M_\odot$, $T_{\rm WD} = 15\times 10^6$ K, and
         $\dot{M} = 2\times 10^{-10}\,M_\odot/\mbox{yr}$. The accreted material was assumed to be a mixture of 50\% solar
         and 50\% WD's core compositions. Our final abundances agree very well with those presented by
         \cite{jh98} in their Fig.~1 for a CO nova model with similar parameters.
         }
\label{fig:f5}
\end{figure}


\begin{figure}
\epsffile [75 250 480 695] {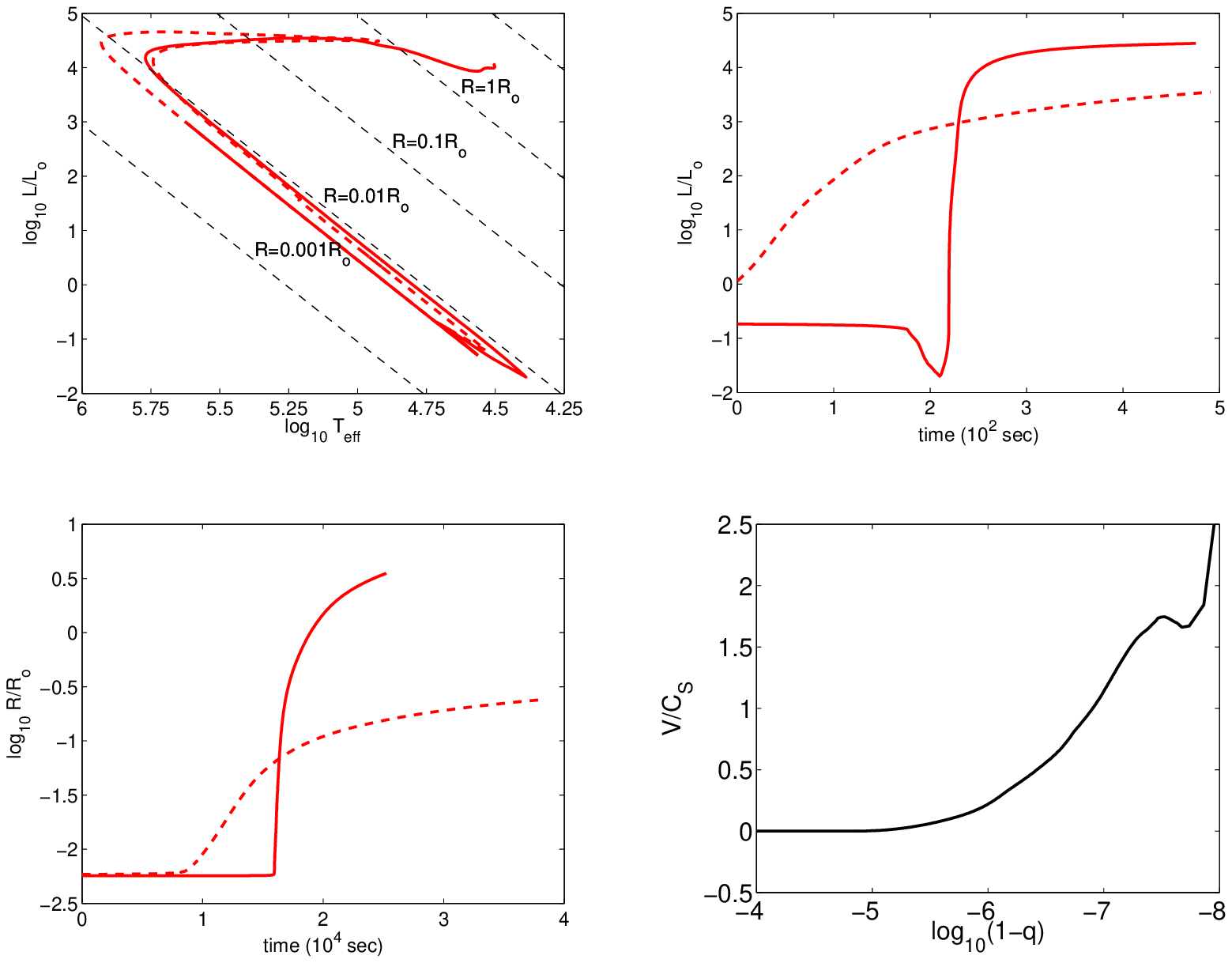}
\caption{Results of the $1.2\,M_\odot$ CO nova simulation with
  convective boundary mixing according to Equation \ref{eq:DOV} (solid red curves). Upper-left
  panel: HRD; upper-right and
  lower-left panels: very fast increase of the surface
  luminosity and a bit slower radial expansion; lower-right panel:
  velocity profile (in the units of the sound speed) in one
  of the last models. For comparison, the dashed red curves show the results for the same
  model but without CBM.}
\label{fig:f6}
\end{figure}


\begin{figure}
\epsfxsize=10cm
\epsffile [60 200 320 695] {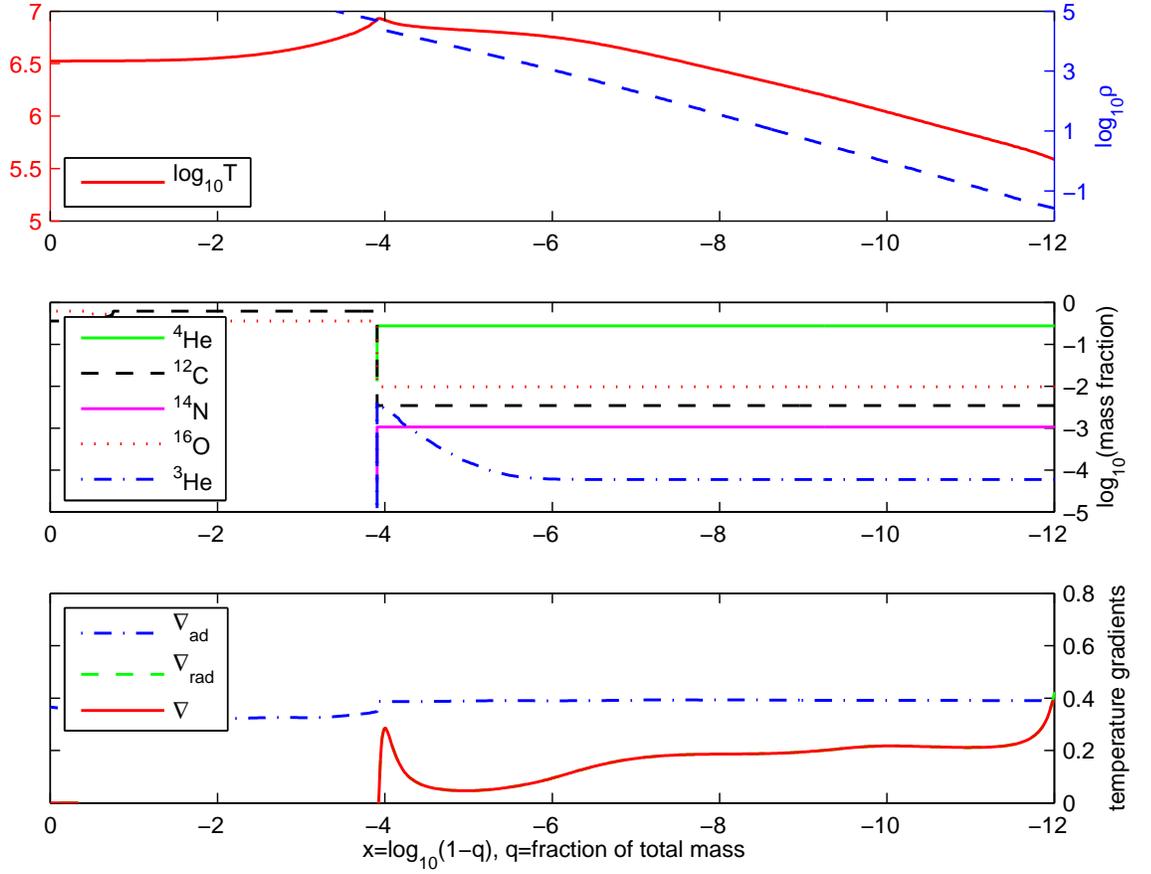}
\caption{Snapshot similar to \abb{fig:f2} for a $1.2\,M_\odot$ CO WD model with the central temperature
  $T_{\rm WD} = \natlog{3.3}{6}\kelv$ that accretes with the rate
  $\dot{M} = 10^{-11}\msun/\jahre$. Note formation of a sloped $^3$He enhancement at
  the base of the accreted envelope (the dot-dashed blue curve in the
  middle panel) during the accretion. The envelope is convectively stable yet, therefore $\nabla$ coincides
  with $\nabla_{\rm rad}$ everywhere (the lower panel).}
\label{fig:f7}
\end{figure}


\begin{figure}
\epsfxsize=10cm
\epsffile [60 200 320 695] {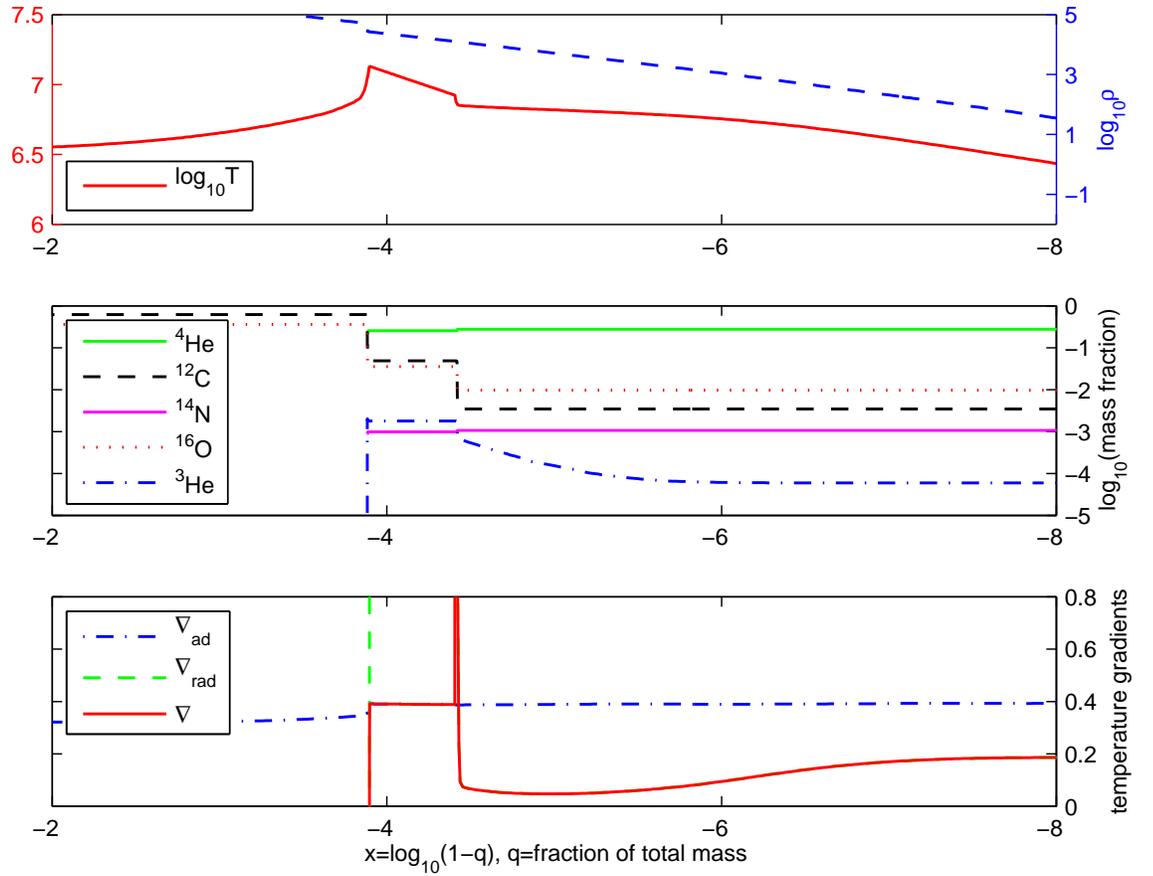}
\caption{Same as \abb{fig:f7}, but at a later time. $^3$He ignition
  triggers a convective zone (the region where $\nabla_{\rm rad} > \nabla \ga \nabla_{\rm ad}$
  in the lower panel). Note the enhanced amount of C and O
  (middle panel) in this model with CBM, sufficient to produce
  a fast CO nova.}
\label{fig:f8}
\end{figure}

\clearpage
\begin{deluxetable}{ccccccc}
  \tablecolumns{7} \tabletypesize{\footnotesize}
  \tablecaption{Characteristics of Our CO Nova Models}
  \tablewidth{0pt} \tablehead{ \colhead{$M_{\rm WD} [\msun]$} &
    \colhead{$T_{\rm WD}[10^6\kelv]$ } & \colhead{$\lg\,L_{WD}[\mathrm{L_\odot}]$} & \colhead{$\lg\,\dot{M}
      [\msun$/yr]} & \colhead{$M_{\rm acc}$ [$10^{-5}\msun$]} &
    \colhead{$\lg\,L_\mathrm{H} [\mathrm{L_\odot}]$} &
    \colhead{$T_{\rm max}[10^6\kelv]$} } 
\startdata
  0.65 & 30 & -1.65 & -9 & 22.0 & 8.06 & 127 \\
  0.65 & 30 & -1.65 & -10 & 22.5 & 8.10 & 126 \\
  0.65 & 30 & -1.65 & -11 & 22.7 & 8.14 & 129 \\
  \tableline
  0.85 & 15 & -2.35 & -9 & 9.8 & 8.58 & 154 \\
  0.85 & 15 & -2.35 & -10 & 12.2 & 9.05 & 165 \\
  0.85 & 15 & -2.35 & -11 & 11.9 & 9.03 & 164 \\
  \tableline
  1.0  & 30 & -1.55 & -9 &  5.3 & 8.96 & 181 \\
  1.0  & 30 & -1.55 & -10 &  5.8 & 9.03 & 184 \\
  1.0  & 30 & -1.55 & -11 &  5.7 & 9.02 & 184 \\
  \tableline
  1.15  & 30 & -1.50 & -10 &  2.6 & 8.87 & 213 \\
  1.15\tablenotemark{a}  & 30 & -1.50 & -10 &  0.6 & 8.63 & 159 \\
  1.15\tablenotemark{b}  & 30 & -1.50 & -10 &  2.6 & 10.67 & 222 \\
  1.15  & 25 & -1.70 & -10 &  2.9 & 8.92 & 218 \\
  1.15  & 20 & -1.94 & -10 &  3.2 & 8.97 & 223 \\
  1.15  & 15 & -2.25 & -10 &  3.6 & 9.03 & 229 \\
  1.15  & 12 & -2.50 & -10 &  3.5 & 8.95 & 222 \\
  1.15  & 10 & -2.69 & -10 &  4.1 & 9.08 & 234 \\
  1.15  &  7 & -3.07 & -10 &  5.8 & 9.26 & 249 \\
  1.15  &  7 & -3.07 & -11 &  5.2 & 9.42 & 244 \\
  \tableline
  1.2  & 30 & -1.49  & -9 &  1.6 & 8.68 & 220 \\
  1.2  & 30 & -1.49 & -10 &  1.9 & 8.74 & 225 \\
  1.2\tablenotemark{a}  & 30 & -1.49 & -10 &  0.3 & 8.47 & 156 \\
  1.2\tablenotemark{b}  & 30 & -1.49 & -10 &  1.9 & 10.67 & 232 \\
  1.2  & 30 & -1.49 & -11 &  1.8 & 8.74 & 225 \\
  \tableline
  1.2  & 20 & -1.92 & -9 &  1.4 & 8.62 & 215 \\
  1.2  & 20 & -1.92 & -10 &  2.4 & 8.86 & 237 \\
  1.2  & 20 & -1.92 & -11 &  2.3 & 8.85 & 238 \\
  \tableline
  1.2  & 15 & -2.23 & -9 &  1.8 & 8.68 & 221 \\
  1.2  & 15 & -2.23 & -10 &  2.2 & 8.76 & 228 \\
  1.2  & 15 & -2.23 & -11 &  2.7 & 8.92 & 242 \\
  \tableline
  1.2  & 7 & -3.04 & -11 & 4.0 & 9.14 & 261 \\
  1.2  & 3.3 & -3.74 & -11 & 14.2 & 9.70 & 328 \\
                      \enddata
                         \label{tab:tab1}
\tablenotetext{a}{This model accretes a mixture of 70\% solar and 30\% WD's compositions.}
\tablenotetext{b}{This simulation includes CBM (\ref{eq:DOV}) with $f=0.004$.}
                         \end{deluxetable}


\end{document}